\documentstyle[12pt,aps,here]{revtex}
\tightenlines

\def\And{{\rm and\ }}
\def\stars{\bigskip\centerline{***}\medskip}
\newif\ifboo \boofalse
\def\Review#1{\boofalse{\it #1},}
\def\Name#1{{\sc #1},}
\def\Vol#1{\ifboo Vol. {\bf #1}\else{\bf #1}\fi}
\def\Year#1{\ifboo #1\else(#1)\fi}
\def\Book#1{\bootrue{\it #1},}
\def\Page#1{\ifboo {\rm p. #1}\else{\rm #1}\fi}
\begin{document}
\title{Copolymer Networks:\\
Multifractal dimension spectra in polymer field theory}
\author{C.~von~Ferber$^{1,2}$ and Yu.~Holovatch$^3$}
\address{
$^1$Fachbereich Physik, Universit\"at - Gesamthochschule - Essen,\\
D-45117 Essen, Germany \\
$^2$School of Physics and Astronomy, Tel Aviv University,
    IL-69978 Tel-Aviv, Israel\\
$^3$Institute for Condensed Matter Physics,
Ukrainian Academy of Sciences,\\
UA-290011 Lviv, Ukraine
}
\maketitle
%
%
%
%
%
\begin{abstract}
We explore the rich scaling behavior of copolymer networks in solution.
We establish a field theoretic description in terms of composite
operators. Our 3rd order resummation of the spectrum of scaling
dimensions brings about remarkable features: Convexity of the spectra
allows for a multifractal interpretation. This has not been conceived
for power of field operators of $\phi^4$ field theory before. 
The 2D limit of the mutually avoiding walk star apparently
corresponds to results of a conformal Kac series. Such a classification
seems not possible for the 2D limit of other copolymer stars.
The 3rd order calculation of a large collection of exponents furthermore
allows for a consistency check of two complementary schemes: epsilon
expansion and renormalization at fixed dimension.
\end{abstract}
\pacs{61.41.+e, 64.60.Ak, 64.60.Fr, 11.10.Gh-}
\section{Introduction}
Recently much interest focused on the relation of field theory and
multifractals \cite{Duplantier91,Ludwig90}
and the associated multifractal dimension spectra
\cite{Cates87,Fourcade87}
as well as
non-intersecting random walks and their 2D conformal theory
\cite{Duplantier88}.
We
present a model of multicomponent polymer networks that shows a
common core
of these topics and allows for a detailed study of the
interrelations. The flux of diffusion onto an absorbing fractal
defines a multifractal measure. Cates and Witten \cite{Cates87} have
mapped the moments of this flux  to that of a star of random walks
(RW) avoiding the absorber taken to be a polymer or RW itself. Using
the field theoretic formulation of polymer theory we show that the
spectrum of scaling exponents governing these problems is given by the
anomalous dimensions of composite operators with appropriate symmetry.

For polymer networks
consisting of polymer chains of one species it has
been shown, that the basic scaling exponents are connected with
'stars', polymer chains tied together at one core 
\cite{Duplantier86,Duplantier89,Schaefer92}.
The number of configurations ${\cal Z}_{*f}$ of a polymer star with
$f$ arms of $N$ monomers will scale for large $N$ like
\begin{equation}\label{0}
{\cal Z}_{*f} \sim N^{\gamma_f-1} \sim (R/\ell)^{\eta_f - f\eta_2}.
\end{equation}
The second part shows scaling with the size $R\sim N^\nu$ of the
isolated coil of $N$ monomers on some scale $\ell$. The
exponents $\nu=3/4,0.58(8)$ and 
$\gamma_1=\gamma_2=\gamma=43/32,1.16(0)$ for space dimensions $d=2,3$
are known in polymer theory \cite{desCloizeaux90}.
The exponents  $\gamma_f$
have been calculated analytically in perturbation theory
\cite{Duplantier89,Schaefer92,Miyake83},
by exact methods in
two dimensions \cite{Duplantier86},
and by Monte Carlo simulations \cite{Batoulis89}.

At short distance two polymer stars will repel each other. In view of
the below advocated language of field theory this is described in
terms of a short distance expansion. One finds the following
relation for the probability $P(r)$ to find the cores of two stars 
of $f_1$ and $f_2$ at short distance $r$ \cite{Duplantier89}
\begin{equation}
 P(r) \sim r^{\Theta} \mbox{ , } 
\Theta=\eta_{f_1} + \eta_{f_2} - \eta_{f_1+f_2} > 0 \;.
\end{equation}
This is compatible with the result, that the spectrum of polymer star
exponents $\eta_f$ is convex from below as function of $f$ with
$\eta_1=0$.

On the other hand a multifractal (MF) measure $\mu_{x}$ defined on
the sites $x$ of scale $\ell $ on
some object of size $R$ is characterized by the scaling of its
moments averaged over all sites:
\begin{equation}
 \langle \mu_{x}^k \rangle = \sum_x \mu_{x}^k 
\sim (R/\ell)^{y_f} \;.
\end{equation}
From general inequalities for the moments of a probabitity distribution 
one may deduce that the spectrum of exponents $y_f$ has to
be convex from above. 
This indicates an apparent discrepancy between objects described in
field theory (FT) as powers of field (see below) such as polymer stars, and
the moments of a MF measure \cite{Duplantier91}. 
This we want to resolve by including both
concepts in the same FT formalism showing that they are special cases
of a more general approach, which in addition also describes the
problem of non-intersecting random walks.  

To this end we study the scaling behavior of a polymer
star or a general {\em network} of chains {\em of different species} and
thus, within a unique formalism, include effects caused by self and
mutual interactions between polymers of different species forming a
network. We combine the field theoretic formalism developed
for the description of polymer stars and networks \cite{Schaefer92}
with the corresponding theory which describes multicomponent
polymer solutions \cite{Schaefer91}.
\section{Theory}
We introduce a
Landau-Ginsburg-Wilson-Lagrangian $\cal L$ of $f$ interacting
fields $\phi_b$ each with $n$ components,
i.e.  $\phi_a^2 = \sum_{\alpha = 1}^{n}( \phi_a^{\alpha} )^2$,
with an interaction matrix $u_{aa'} $ and mass parameters $m_a$:
\begin{eqnarray}
\label{4}
{\cal L}\{\phi_b,m_b\} &=& \frac{1}{2} \sum_{a=1}^{f}\int{\rm d}^d r
\left(m_a\phi_a^2 + (\nabla \phi_a(r) )^2 \right)
+ \frac{1}{4!} \sum_{a,a^{'}=1}^{f} u_{aa'}
\int {\rm d}^d r \phi_a^2(r)\phi_{a'}^2(r) .
\end{eqnarray}
In this theory the star exponents are given in terms of the anomalous
dimensions of composite operators $\prod_{a=1}^{f}\phi_a$
\cite{Schaefer92}. We define vertex functions $\Gamma^{*f} $
with insertion of this operator by
\begin{equation}
\label{7}
\delta(q_0 + \ldots + q_f)
\Gamma^{*f}(q_0 \dots q_f) =
\int \prod_{k=0}^{f} e^{i(q_k r_k)} {\rm d}^d r_k
\langle \prod_{a=1}^{f} \phi_{a}(r_0)
\phi_{1}(r_1)\dots \phi_{f}(r_f)
\rangle^{\cal L}_{{\rm 1pi},n=0},
\end{equation}
As in standard polymer FT
this is evaluated with respect to the Lagrangian
(\ref{4}) keeping only  contributions which correspond
to one particle irreducible (1pi) graphs which have nonvanishing
tensor factors in the $n=0$ limit.
In the single component case the theory may also be described
in terms of one $O(n)$ symmetric field $\phi$ with $n>f$, where the
corresponding operator is
$N^{\alpha_1\ldots \alpha_f}\phi^{\alpha_1}\cdots\phi^{\alpha_f}$
with a traceless tensor $N^{\alpha_1\ldots \alpha_f}$ in the formal
limit $n=0$ \cite{Schaefer92,Wallace75}.

We apply RG theory to make use
of the scaling symmetry of the systems in the asymptotic limit to
extract the universal content and at the same time remove divergences
which occur for the evaluation of the bare functions in this limit
\cite{Brezin76}.
Several asymptotically equivalent
procedures serve to the purpose of renormalization. 
In the present study we use two somewhat complementary
approaches: zero mass renormalization with successive $\varepsilon=4-d$
-expansion \cite{Brezin76}
and the massive RG approach at fixed dimension
\cite{Parisi80}. Application of both approaches will enable us 
to check the consistency of approximations and the accuracy
of the results obtained.
We pass from the theory in terms of the initial bare variables to a
renormalized theory. This can be achieved by a controlled rearrangement of
the series for the vertex functions (\ref{7}) introducing
renormalizing $Z$-factors for fields ($Z_{\phi_a}$), couplings
($Z_{ab}$) and mass. Then, for instance the bare couplings
$u_{ab}$ are given in terms of their renormalized dimensionless counterparts
$g_{ab}$ by
\begin{equation}
\label{10}
u_{ab} = \kappa^{4-d} Z_{\phi_a}Z_{\phi_b}Z_{ab} g_{ab} \;.
\end{equation}
The scale parameter $\kappa$ represents the mass at which the
massive scheme is evaluated and the scale of external momenta
in the massless $\varepsilon$-expansion scheme.
We define the $Z$-factors in (\ref{10})
as to renormalize the correlators $\langle\cdots\rangle^{\cal L}$ 
in each RG procedure (see e.g.\cite{Brezin76}). 
The polymer limit $n=0$ of zero
component fields leads to essential simplification. Each field
$\phi_a$, mass $m_a$ and coupling $u_{aa}$ renormalizes as if the
other fields were absent. The renormalization of the couplings
$u_{ab}$ involves only the fields $\phi_a$,$\phi_b$ \cite{Schaefer91}.
The renormalized couplings $g_{ab}$ defined by relations
(\ref{10}) depend on the scale parameter $\kappa$.
Thus the renormalization $Z$ - factors also
depend implicitly on $\kappa$. This dependence
defines the RG functions and exponents:
$\kappa \frac{\rm d}{{\rm d}\kappa} g_{aa} = \beta_{aa}(g_{aa})$;
$\kappa \frac{\rm d}{{\rm d}\kappa} g_{ab} = \beta_{ab}(g_{aa},g_{bb},g_{ab})$;
$\kappa \frac{\rm d}{{\rm d}\kappa} \ln Z_{\phi_a} = \eta_{\phi_a}(g_{aa})$.
The function $\eta_{\phi_a}$ defines the pair correlation critical exponent.
The set of scaling exponents $\eta_{*f}$ for general copolymer stars is
defined by the renormalization factors $Z_{* f}$ for the 
star vertex functions $\Gamma^{*f}$:
\begin{equation}
\label{11}
\prod_{a=1}^{f} Z_{\phi_a}^{1/2}
Z_{* f}
\Gamma^{* f}(u_{bb'}(g_{bb},g_{b'b'},g_{bb'})) = \kappa^{\delta_{f}},
\mbox{ with }\;
\eta_{*f}(g_{ab})=\kappa \frac{\rm d}{{\rm d}\kappa} \ln Z_{* f}\;.
\end{equation}
$\delta_f = d+(1-d/2)f$
is the engineering dimension of the
corresponding bare vertex function.

In a study devoted to ternary polymer solutions the RG flow
given by the above defined $\beta$-functions
has been calculated \cite{Schaefer91,FerHol}
to third loop order. The equations for the fixed
points of the $\beta$-functions were found to
have the following nontrivial solutions:
$\beta_{{aa}}(g^*_{\rm S})= 0$
and for $a\neq b $:
$\beta_{{ab}}(0          ,0          , g^*_{\rm G}) = 0$,
$\beta_{{ab}}(g^*_{\rm S},0          , g^*_{\rm U}) = 0$,
$\beta_{{ab}}(0          ,g^*_{\rm S}, g^*_{\rm U}) = 0$,
$\beta_{{ab}}(g^*_{\rm S},g^*_{\rm S}, g^*_{\rm S}) = 0$,
 corresponding to all combinations of
interacting and non-in\-ter\-ac\-ting chains. 

We evaluate the exponents for two general arrangements of the fixed
point matrix. The ternary case of two mutually interacting species of
polymer chains in solution, and the mutual avoiding walk case of
essentially $f$ {\em only mutually} interacting species. In the first case
we describe polymer stars made of $f_1$ chains of species 1 and $f_2=f-f_1$
chains of species 2. Either both species are non self-interacting
and
\begin{equation}
\eta^G_{f_1f_2} \equiv 
\eta_{*f}(g_{ab}=0 \mbox{ if } a,b\leq f_1 \mbox{ or } a,b>f_1; 
                        \mbox{ else } g_{ab}=g^*_{\rm G})\;,
\end{equation}
or species 1 self-interacts and species 2 does not such that
\begin{equation}
\eta^U_{f_1f_2} \equiv 
\eta_{*f}(g_{ab}=g^*_{\rm S} \mbox{ if } a,b\leq f_1;
     g_{ab}=0 \mbox{ if } a,b>f_1; \mbox{ else } g_{ab}=g^*_{\rm U}).
\end{equation}
For $f_2=0$ this includes the homo-polymer star with
$\eta_f=\eta^U_{f,0}$ in eq.(\ref{0}).
The mutually avoiding walk case reads
\begin{equation}
\eta^{{\rm MAW}}_f \equiv \eta_{*f}(g_{ab}=0 \mbox{ if } a=b 
\mbox{ else } g_{ab}=g^*_{\rm G})\;.
\end{equation}
\section{Results}
We give the results for the exponents in $\varepsilon=4-d$-expansion.
The corresponding more lengthy expressions  obtained by fixed $d=3$ RG
may be found in \cite{FerHol}:
\begin{equation}
\eta^G_{f_1f_2} (\varepsilon) =
-{{\it f_1}\,{\it f_2}\frac{\varepsilon}{2}}+
{{\it f_1}\,{\it f_2}\,\Big ({\it f_2}-3+
{\it f_1}\Big)\frac{{\varepsilon}^{2}}{8}}-
{{\it f_1}\,{\it f_2}\,\Big ({\it f_2}-3+
{\it f_1}\Big )\Big ({ \it f_1}+{\it f_2}+
3\,\zeta (3)-
3\Big)\frac{{\varepsilon}^{3}}{16}}
\label{18}
\end{equation}
\begin{eqnarray}
\arraycolsep0pt
\lefteqn{
\eta^U_{f_1f_2}(\varepsilon) =
{\it f_1}\,\Big (1-{\it f_1}-
3\,{\it f_2}\Big )\frac {\varepsilon}{8} +
{\it f_1}\,\Big (25-33\,{\it f_1}+
8\,{{\it f_1}}^{2}-91\,{\it f_2}+
42\,{\it f_1}\,{\it f_2}+
18\,{{\it f_2}}^{2}\Big) \frac {\varepsilon^2}{256}}
\nonumber \\ &&
+ {\it f_1}\,\Big (577
-969\,{\it f_1}+
456\,{{\it f_1}}^{2}-
64\,{{\it f_1}}^{3}-2463\,{\it f_2}+
2290\,{\it f_1}\,{\it f_2}-
492\,{{\it f_1}}^{2}{\it f_2}+
1050\,{{\it f_2}}^{2}
\nonumber \\ &&
-504\,{\it f_1}\,{{\it f_2}}^{2}-
108\,{{\it f_2}}^{3}-
712\,\zeta (3)+
936\,{\it f_1}\,\zeta (3) -
224\,{{\it f_1}}^{2}\zeta (3)
\nonumber \\  &&
+2652\,{\it f_2}\,\zeta (3)
-1188\,{\it f_1}\,{\it f_2}\,\zeta (3)-
540\,{{\it f_2}}^{2}\zeta (3)\Big )
\frac {\varepsilon^3}{4096}
\label{19}
\end{eqnarray}
\begin{equation}
\eta^{{\rm MAW}}_{f}(\varepsilon) =
-\Big ({\it f}-
1\Big ){\it f}\, \frac {\varepsilon}{4}+
{\it f}\,\Big ({\it f}-
1\Big )\Big (2\,{\it f}-
5\Big ) \frac {\varepsilon^2}{16}-
\Big ({\it f}-
1\Big ){\it f}\,\Big (4\,{{\it f}}^{2}-
20\,{\it f}+8\,{\it f}\,\zeta (3)-
19\,\zeta (3)+
25\Big )
\frac {\varepsilon^3}{32}
\label{20}
\end{equation}
Here $\zeta(3)\simeq 1.202$ is the Riemann $\zeta$-function.
The above formulas reproduce the 3rd order calculations of 
$\gamma_f-1=\nu(\eta^{\rm U}_{f,0}-f\eta^{\rm U}_{2,0})$
\cite{Schaefer92} as well as the 2nd order 
exponents $\lambda^{\rm (xx)}$ defined in equations ${\rm (xx)}$ of
\cite{Cates87}, $\lambda^{(29)}(n)=-\eta^{\rm G}_{2,n}$,
$\lambda^{(47)}(n)=-\eta^{\rm U}_{2,n}+\eta^{\rm U}_{2,0}$,
$\lambda^{(48)}_{\rm e}(n)=-\eta^{\rm G}_{1,n}$,
$\lambda^{(49)}_{\rm e}(n)=-\eta^{\rm U}_{1,n}$,
correcting a missprint in eq.(49) of \cite{Cates87}.
Also the 2nd order results for exponents 
$x_{L,n}-x_{L,1}=-2(\eta^{\rm G}_{L,n} - \eta^{\rm G}_{L,1}) $
of \cite{Duplantier91} and 
$\sigma_L=1/2\eta^{\rm MAW}_L $ defined in \cite{Duplantier88}
find their 3rd order extension by the above expansions.

With these exponents we can describe the scaling behavior of
polymer stars and networks of two components, generalizing
the relation for single component networks \cite{Duplantier89}.
In the notation of (\ref{0})
we find for the number of configurations of a network $\cal G$
of $F_1$ and $F_2$ chains of species $1$ and $2$
\begin{equation}
\label{20a}
{\cal Z}_{\cal G} \sim
(R/\ell)^{\eta_{\cal G} -F_1\eta_{20}-F_2\eta_{02}} \mbox{,with }
\eta_{\cal G} = -d L + \sum_{f_1+f_2\geq 1} N_{f_1f_2}\eta_{f_1f_2},
\end{equation}
where $L$ is the number of Loops and $N_{f_1f_2}$ the number of
vertices with $f_1$ and $f_2$ arms of species $1$ and $2$ in
the network $\cal G$. To receive an appropriate scaling law
we assume the network to be built of chains which for both species
will have a coil radius $R$ when isolated.

To obtain reliable numerical values from the  $\varepsilon$-expansions 
in (\ref{18}) - (\ref{20}) and from the series
obtained in the fixed $d$ scheme \cite{FerHol} we apply
Borel resummation using the technique of conformal mapping 
\cite{Leguillou80} which has proven to yield good results for many
critical exponents. We use information
about the higher order behavior \cite{Leguillou80,Schaefer91}
 of the series
(\ref{18})-(\ref{20}) derived from the instanton analysis of the
appropriate field theory.  The results for $d=3$ are given in
Table \ref{tab1}. 
The data show consistency and stability of the results while deviations
grow for large number of arms as may be expected. Note that the 
above expansions are in fact series in $f\varepsilon$, not 
$\varepsilon$ alone.
\section{Conclusions}
\subsection{Multifractals and Field Theory}
Does the data answer the question of convexity?
A close study of the matrix of values reveals, that for fixed $f_1$
both $\eta^{\rm G}_{f_1f_2}$ and $\eta^{\rm U}_{f_1f_2} $
are convex from above as function of $f_2$, thus yielding `MF
statistics'. The relation to a MF spectral function for $f_1=1,2$ has
been pointed out in \cite{Cates87}, it is analysed in close detail in
view of the new data and FT formulation in a separate publication 
\cite{FerHol}.
On the other hand also copolymer stars should repel each other.
This is found to be true as well, the corresponding convexity from
below shows up e.g. along the diagonal values $\eta_{ff}$ as
function of $f$. The general relation 
$\eta_{f_1f_2}+\eta_{f'_1f'_2} \geq \eta_{f_1+f'_1,f_2+f'_2} $
is always fulfilled. In view of our FT formalism the MF moments
$\langle\mu^k \rangle$ are represented by field operators 
$\phi_a^L\phi_b^k=\phi_{a_1}\cdots\phi_{a_L}\phi_{b_1}\cdots\phi_{b_k}$
in a FT with vanishing interactions $g_{b_ib_j}$.
Thus, even though simple power $k$ of field operators $\phi^k$ do not 
describe MF moments \cite{Duplantier91},
they may be written as a power $L+k$ of field operators which
have the appropriate short distance behavior. This is also illustrated
in fig.1, showing the spectrum of exponents $\eta^{\rm U}_{f_1f_2}$ in
the 2D limit \cite{FerHol}. The opposite convexity along the two
axes is clearly seen for these unsymmetric combinations of a polymer
$f_1$-star and a random walk $f_2$-star which mutually interact. 
\subsection{2D Copolymer Stars}
The 2D exponents for polymer stars have been shown
to belong to a Kac series of exponents of conformal FT with 
$\gamma_f-1=(4+27f-9f^2)/64$ \cite{Duplantier86}.
There are strong indications that this is the case also for 
MAW stars with $\eta^{\rm MAW}_f= (1-4f^2)/12$ \cite{Duplantier88}.
Already in view of fig.1 though, such a simple 2nd order polynomial
seems not to describe the 2D limit of general copolymer star
exponents. In 2D however, each chain of a star will interact only with
its direct neighbors. A star described here by $\eta^{\rm G}_{ff} $
will behave like a MAW $2f$-star if each species-1 chain has two
neighbors of species-2 whereas it will behave differently if the chains
are ordered such that each species is in one bulk of chains. The 2D
copolymer stars in this sense reveal an even richer behavior.
Thus, the copolymer generalization of the MAW star adds another problem, 
for which a rigorous formulation in terms of an exactly solvable 2D
model is yet to be found. 
\stars\\
We thank Lothar Sch\"afer for valuable discussions.
Supported in part by Deutsche Forschungsgemeinschaft SFB 237
and Minerva Gesellschaft.

\newpage
\begin{table}[H]
\caption{ \label{tab1}
Values of the copolymer star exponent $\eta_{f_1f_2}^{U}$, upper
part {\rm(U)}, and  $\eta_{f_1f_2}^{G}$, lower part {\rm(G)},
at $d=3$ obtained by $\varepsilon$-expansion
($\varepsilon$) and by fixed dimension technique
($3d$).
}
\tabcolsep1.4mm
\begin{tabular}{llrrrrrrrrrrrr}
&$f_2$ &
\multicolumn{2}{c}{$1$}&
\multicolumn{2}{c}{$2$}&
\multicolumn{2}{c}{$3$}&
\multicolumn{2}{c}{$4$}&
\multicolumn{2}{c}{$5$}&
\multicolumn{2}{c}{$6$}\\
&$f_1$ & 
$\varepsilon$ & $3d$ &
$\varepsilon$ & $3d$ &
$\varepsilon$ & $3d$ &
$\varepsilon$ & $3d$ &
$\varepsilon$ & $3d$ &
$\varepsilon$ & $3d$ \\
\hline
&1 & 
  -0.43 &  -0.45 &
  -0.79 &  -0.81 &
  -1.09 &  -1.09 &
  -1.35 &  -1.37 &
  -1.60 &  -1.64 &
  -1.81 &  -1.89\\
&2 &
  -0.98 &  -0.98 &
  -1.58 &  -1.60 &
  -2.13 &  -2.19 &
  -2.61 &  -2.71 &
  -3.05 &  -3.21 &
  -3.46 &  -3.68\\
U&3 &
  -1.64 &  -1.67 &
  -2.44 &  -2.52 &
  -3.16 &  -3.30 &
  -3.82 &  -4.04 &
  -4.44 &  -4.75 &
  -5.01 &  -5.42\\
&4 &
 -2.39 &  -2.47 &
 -3.33 &  -3.50 &
 -4.20 &  -4.48 &
 -5.02 &  -5.40 &
 -5.80 &  -6.30 &
 -6.53 &  -7.15\\
&5 &
 -3.21 &  -3.38 &
 -4.28 &  -4.57 &
 -5.28 &  -5.71 &
 -6.24 &  -6.81 &
 -7.15 &  -7.89 &
 -8.02 &  -8.92\\
&6 &
 -4.11 &  -4.40 &
 -5.29 &  -5.73 &
 -6.41 &  -7.03 &
 -7.48 &  -8.28 &
 -8.51 &  -9.50 &
 -9.50 & -10.69\\
\hline
&1 &  
  -0.56 &  -0.58 &
  -1.00 &  -1.00 &
  -1.33 &  -1.35 &
  -1.63 &  -1.69 &
  -1.88 &  -1.98 &
  -2.10 &  -2.24 \\
&2 & & &
  -1.77 &  -1.81 &
  -2.45 &  -2.53 &
  -3.01 &  -3.17 &
  -3.51 &  -3.75 &
  -3.95 &  -4.28 \\
G&3 & & & & &
 -3.38 &  -3.57 &
 -4.21 &  -4.50 &
 -4.94 &  -5.36 &
 -5.62 &  -6.15 \\
&4 & & & & & & &
 -5.27 &  -5.71 &
 -6.24 &  -6.84 &
 -7.12 &  -7.90 \\
&5 & & & & & & & & &
 -7.42 &  -8.24 &
 -8.50 &  -9.54 \\
&6 & & & & & & & & & & &
 -9.78 &  -11.07
\end{tabular}
\end{table}
\begin{figure}[H]
\input{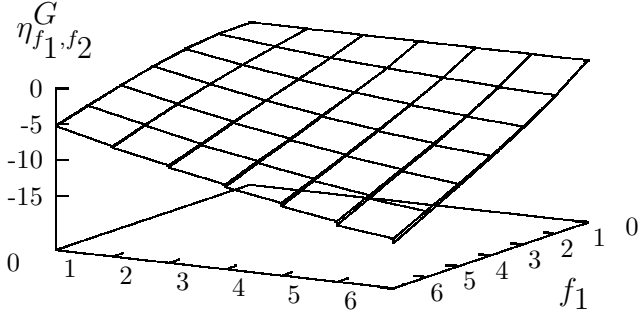}
\caption{Exponent $\eta^{\rm U}_{f_1f_2}$ in the `Unsymmetric' fixed point
at $d=2$ obtained in $\epsilon$-expansion and in fixed $d$
scheme. The steps in the `flying carpet' indicate the difference of
the results in the two approaches}
\label{fig1}
\end{figure}
\end{document}